\documentclass[]{iopart}
\usepackage{graphicx}
\usepackage{dcolumn}
\usepackage{amssymb}
\expandafter\let\csname equation*\endcsname\relax
\expandafter\let\csname endequation*\endcsname\relax
\usepackage{amsmath}
\usepackage{amstext}
\usepackage{amsthm}
\usepackage{latexsym}
\usepackage{verbatim}
\usepackage{color} 
\usepackage{multirow}
\usepackage{clrscode}
\usepackage{hyperref}
\usepackage{braket}
\usepackage{array}

\newcommand{\var}{\mathop{\mathrm{var}}}

\usepackage{fullpage}

\begin{document}

\title {The Ising chain constrained to an even or odd number of positive
  spins 
}
\author{Michael T.~Gastner$^{1,2}$}
\address{$^1$Institute of Technical Physics and Materials Science,
  Research Centre for Natural Sciences, Hungarian Academy of Sciences,
  P.O. Box 49, H-1525 Budapest, Hungary}
\address{$^2$Department of Engineering Mathematics, University of
  Bristol, Merchant Venturers Building, Woodland Road, Bristol BS8
  1UB, United Kingdom.}
\ead{m.gastner@bristol.ac.uk}
\begin{abstract}
  We investigate the statistical mechanics of the periodic
  one-dimensional Ising chain when the number of positive spins is
  constrained to be either an even or an odd number.
  We calculate the partition function using a generalization of the
  transfer matrix method.
  On this basis, we derive the exact magnetization, susceptibility,
  internal energy, heat capacity and correlation function.
  We show that in general the constraints substantially slow down
  convergence to the thermodynamic limit.
  By taking the thermodynamic limit together with the limit of zero
  temperature and zero magnetic field, the constraints lead to new
  scaling functions and different probability distributions for the
  magnetization.
  We demonstrate how these results solve a stochastic version of the
  one-dimensional voter model.\\

\end{abstract}

\section{Introduction}
For almost one century, the Ising model of ferromagnetism has been
a cornerstone of statistical mechanics~\cite{Ising25}.
It is one of very few problems that can, at least in one and two
dimensions, be solved exactly~\cite{Baxter82}.
Its applications range from solid state physics~\cite{GrossoPastori00}
over neuroscience~\cite{Schneidman_etal06} to collective social
phenomena~\cite{Stauffer08}.
In its basic form the Ising model is based on the Hamiltonian
\begin{eqnarray}
  E(\boldsymbol{\sigma}) = -J\sum_{i=1}^N{\sigma_i\sigma_{i+1}} - H\sum_{i=1}^N\sigma_i,
  \label{Hamiltonian}
\end{eqnarray}
where each spin in the vector $\boldsymbol{\sigma} =
(\sigma_1,\ldots,\sigma_N)$ can take only the values $\pm1$ and we
assume periodic boundary conditions so that $\sigma_{N+1} = \sigma_1$.
The parameter $J$ is the strength of interactions between spins and
$H$ an external magnetic field.
We allow $J$ to take positive and negative values, thereby considering
both the ferro- and antiferromagnetic case.

Many generalizations of the model have been investigated since Ising's
groundbreaking publication, for example long-range
interactions~\cite{SiegertVezzetti68}, spin
glasses~\cite{SherringtonKirkpatrick75} and permitting more than two
possible spin states~\cite{AshkinTeller43}.
In this article we investigate two different variations of the Ising
model.
First, we restrict the number of positive spins
\begin{eqnarray}
  N_+(\boldsymbol{\sigma}) = \frac12\left(N+\sum_{i=1}^N\sigma_i\right)
\end{eqnarray}
to an even number.
That is, the Hamiltonian is given by Eq.~\ref{Hamiltonian} if $N_+$ is
even and $E=\infty$ if it is odd.
In the second model, Eq.~\ref{Hamiltonian} holds if $N_+$ is odd,
whereas an even $N_+$ is forbidden.
We will refer to these two models as ``even'' or ``odd'' Ising model
respectively~\footnote{Sometimes the term ``odd Ising model'' is used
  for a spin glass model by Villain~\cite{Villain77} which is
  unrelated to our work.}.

In Sec.~\ref{motivation}--\ref{asynch_section} we will motivate
the even and odd model by showing that they are equivalent to a simple
opinion formation model.
In Sec.~\ref{partition_function} we demonstrate how the transfer
matrix method for the unconstrained Ising model can be modified to
derive the partition functions of the even and odd model.
Section~\ref{magn_susc} contains a derivation of the magnetization and
susceptibility of both models.
We deduce the nearest-neighbour correlations, internal energy and heat
capacity in Sec.~\ref{corr_inten_heatcap} and the correlation function
in Sec.~\ref{corrfunc_section}.
As we show in Sec.~\ref{leading_order_correction} and \ref{pdf_section},
the constrained models approach the thermodynamic limit in a different
manner than the usual unconstrained model when the temperature and
magnetic field simultaneously go to zero. 
We apply these results to the opinion formation model in
Sec.~\ref{discussion} before summarizing the key findings in
Sec.~\ref{conclusion}.

Before proceeding, we emphasize that $N_+$ is not a fixed number,
neither in the even nor odd model. It is still
permitted to take a multitude of values (e.g.\ in the even model
$N_+=0,2,4,\ldots,2\lfloor N/2\rfloor$), but with the restriction
that configurations with either odd or even $N_+$ are excluded. 
In a Monte Carlo simulation, this restriction could be imposed by
initializing the spins with an even or odd $N_+$ and subsequently
flipping two distinct spins simultaneously in each update.
Because such a Markov chain is not ergodically exploring the
configurations of the conventional (i.e.\ unconstrained) Ising model,
we should not expect that the equilibrium properties are equal.
One purpose of this article is to convince ourselves that the
thermodynamic limits (i.e.\ $N\to\infty$) of the even and odd models
are indeed the limits of the unconstrained model for fixed temperature.
However, we will point out differences when the thermodynamic limit is
taken simultaneously with the limit of zero temperature and zero
magnetic field.

\section{Motivation: Stochastic synchronous voter model}
\label{motivation}
We consider a version of the voter model with stochastic opinion
updates.
Individuals are placed on the $N$ sites of a one-dimensional chain
with periodic boundary conditions.
Each individual holds one of two possible opinions: ``black'' or
``white''.
We associate each site $i$ with a binary variable $\omega_i$ whose
values are
\begin{eqnarray}
  \omega_i(t) =
  \begin{cases}
    1 & \text{if $i$ is black at time $t$,}\\
    -1 & \text{if $i$ is white at time $t$.}
  \end{cases}
\end{eqnarray}
At each discrete time step $t$, all individuals synchronously update
their opinions~\cite{Gracia_etal13}.
(We will discuss asynchronous updates in Sec.~\ref{asynch_section}.)
Each individual randomly chooses one of their two nearest neighbours
and adopts her opinion with probability $p_+$ or chooses the opposite
opinion with probability $p_-=1-p_+$.
Thus, the probability that $i$'s next opinion is $\Omega=\pm1$ can be
expressed as
\begin{eqnarray}
  \Pr\left[\omega_i(t+1)=\Omega\,|\,\omega_{i-1}(t),\omega_{i+1}(t)\right]
  = 
  \begin{cases}
    p_+&\text{ if
      $\frac\Omega2\left(\omega_{i-1}(t)+\omega_{i+1}(t)\right) =
      1$},\\
    \frac12&\text{ if
      $\omega_{i-1}(t)+\omega_{i+1}(t) =
      0$},\\
    p_-&\text{ if
      $\frac\Omega2\left(\omega_{i-1}(t)+\omega_{i+1}(t)\right) =
      -1$},
  \end{cases}
  \label{voter_model_Pr}
\end{eqnarray}
where the subindices are interpreted modulo $N$ to satisfy the
periodic boundary conditions.

What are the equilibrium properties of this model? For example, how
many pairs of neighbours will on average disagree?
And what are the typical fluctuations around this average value?
We will demonstrate that these questions can be analytically answered
by mapping the problem to an Ising model on the dual lattice with an
even number of negative spins. (We will explain the origin of the
even-numbered constraint in Sec.~\ref{map_voter_to_Ising}.)
For even (odd) $N$, the opinion model will consequently map onto the
even (odd) Ising model.

\begin{figure}
  \begin{center}
    \includegraphics[width=6cm]{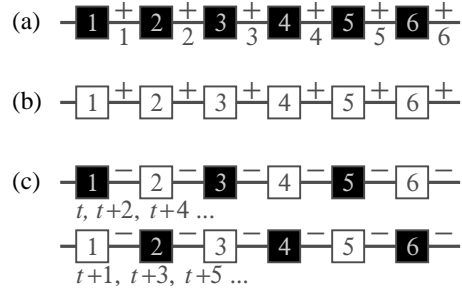}
    \caption{
      The (a), (b) stationary and (c) periodic states of the
      opinion
      dynamics of Eq.~\ref{voter_model_Pr} in the limit $p_+=1$
      with synchronous updates.
      A site $i$ with $\omega_i=\pm1$ is represented by a black
      (white) square.
      The spins of the associated Ising model are shown as $+$ or $-$
      signs above the links.
    }
    \label{voter_stat}
  \end{center}
\end{figure}
Let us first clarify that the variables $\omega_i$ cannot directly be
interpreted as Ising spins $\sigma_i$.
For simplicity's sake, let us assume for a moment that $N$ is even.
In the limiting case of $p_+=1$, there are two stationary states where
all sites have reached either a black or white consensus
(Fig.~\ref{voter_stat}a and~\ref{voter_stat}b).
For synchronous updates there is, however, also a periodic state where
the opinions alternate in space~\cite{Skorupa_etal12}:
if all odd sites are black and all even sites white at time $t$, all
opinions are inverted at $t+1$ and return to the original state at
$t+2$ (Fig.~\ref{voter_stat}c).
Unlike in the zero-temperature Ising model, we thus have apparently
more than two ground states.

We can, however, establish a connection to the Ising model if we
assign spins $\sigma_i$ to the $i$-th link (i.e.\ between the sites
$i$ and $i+1$) rather than the sites themselves.
We set $\sigma_i=1$ if both sites connected by the link agree and
$\sigma_i=-1$ if they disagree,
\begin{eqnarray}
  \sigma_i = \omega_i\omega_{i+1}.
  \label{voter_sigma}
\end{eqnarray}
In terms of $\sigma_i$, both consensus states are mapped to maximally
positive magnetization, whereas in the alternating state all spins are
negative.
Thus, the limit $p_+=1$ can be mapped to the zero-temperature
ferromagnetic Ising model. We will now argue that for any
$0<p_+<1$, there is a finite-temperature Ising model whose equilibrium
properties are those of the original opinion dynamics given by
Eq.~\ref{voter_model_Pr}.

\section{Mapping the synchronous voter model to an odd or even Ising
  model}
\label{map_voter_to_Ising}

Suppose that the opinions at time $t$ are $\omega_1^{(A)}, \ldots,
\omega_N^{(A)}$. What is the probability $\Pr(A\to B)$ to find the
opinions $\omega_1^{(B)}, \ldots, \omega_N^{(B)}$ at time $t+1$?
Assuming that the probabilities in Eq.~\ref{voter_model_Pr} are
independent for all $i$,
\begin{eqnarray}
  \Pr(A\to B) = \prod_{i=1}^N\Pr\left[\omega_i^{(B)} \bigg|
    \omega_{i-1}^{(A)},\,\omega_{i+1}^{(A)}\right].
  \label{PrAtoB}
\end{eqnarray}
We want to show that
\begin{eqnarray}
  \frac{\Pr(A\to B)}{\Pr(B\to A)} =
  e^{\beta(E^{(A)}-E^{(B)})},
  \label{det_bal}
\end{eqnarray}
where $E^{(A)}$ is the energy of the spins
$\sigma_i^{(A)}=\omega_i^{(A)}\omega_{i+1}^{(A)}$ in the Ising model
without magnetic field,
\begin{eqnarray}
  E^{(A)} = -J\sum_{i=1}^N\sigma_i^{(A)}\sigma_{i+1}^{(A)},
\end{eqnarray}
and similarly for state $B$.
Furthermore,
\begin{eqnarray}
  \beta = -\frac{\ln\left(2\sqrt{p_+p_-}\right)}{2J}
  \label{voter_beta}
\end{eqnarray}
so that every $p_+$ can be mapped to a temperature
$\left(k_B\beta\right)^{-1}$, where $k_B$ is the Boltzmann constant.
Equation~\ref{det_bal} is the detailed balance condition for the
Ising model~\cite{NewmanBarkema99}.
Consequently, the equilibrium properties of the spins $\sigma_i$ can
be deduced from the model's partition function.

Before deriving Eq.~\ref{det_bal}, we emphasize that not all spin
configurations are possible.
The number of negative spins must be even; otherwise the opinions
$\omega_i$ would change an odd number of times as we go once through
the chain so that we would not end up with the same opinion with which
we started.
The restriction to an even number of negative spins changes the
partition function of this model compared to the unconstrained Ising
model.

First, however, we still need to justify Eq.~\ref{det_bal}.
Let us denote the number of neighbouring spins with opposite signs in
states $A$ and $B$ by $n^{(A)}$ and $n^{(B)}$ respectively.
Because $E^{(A)}=J\left(2n^{(A)}-N\right)$ and
$E^{(B)}=J\left(2n^{(B)}-N\right)$, we can rewrite the right-hand side
of Eq.~\ref{det_bal} as
\begin{eqnarray}
  e^{\beta(E^{(A)}-E^{(B)})} =
  e^{2\beta J(n^{(A)}-n^{(B)})}.
  \label{rhs_proof}
\end{eqnarray}
Because of Eq.~\ref{voter_model_Pr} and~\ref{PrAtoB}, only factors
$p_+$, $\frac12$ and $p_-$ can appear in $\Pr(A\to B)$ and $\Pr(B\to
A)$,
\begin{eqnarray}
  &\Pr(A\to B) = p_+^{a_1}p_-^{a_2}/2^{a_3},
  \label{a1a2a3}\\
  &\Pr(B\to A) = p_+^{b_1}p_-^{b_2}/2^{b_3}.
  \label{b1b2b3}
\end{eqnarray}
Assuming $p_+\neq\frac12$, the exponents $a_i$, $b_i$ are uniquely
determined.
(If $p_+=\frac12$ and thus $\beta=0$, Eq.~\ref{det_bal} is trivially
correct.)
There is one factor for each site, so
\begin{eqnarray}
  a_1+a_2+a_3 = b_1+b_2+b_3 = N.
  \label{sum_ai}
\end{eqnarray}
Because $\sigma_i\sigma_{i+1}=-1$ if and only if
$\omega_i+\omega_{i+2} = 0$, Eq.~\ref{voter_model_Pr} implies
\begin{eqnarray}
  a_3 = n^{(A)},
  \label{a3}\\
  b_3 = n^{(B)}.
  \label{b3}
\end{eqnarray}
From Eq.~\ref{voter_model_Pr} it also follows that
\begin{eqnarray}
  &a_1-a_2 =
  \frac12\sum_i\omega_i^{(B)}
  \left(\omega_{i-1}^{(A)}+\omega_{i+1}^{(A)}\right),
  \label{a1_minus_a2}\\
  &b_1-b_2 = \frac12\sum_i\omega_i^{(A)}
  \left(\omega_{i-1}^{(B)}+\omega_{i+1}^{(B)}\right).
  \label{b1_minus_b2}
\end{eqnarray}
Splitting the sums and shifting the summation index shows that
the sums in Eq.~\ref{a1_minus_a2} and~\ref{b1_minus_b2} are equal,
thus
\begin{eqnarray}
  a_1-a_2 = b_1-b_2.
  \label{a1_minus_a2=b1_minus_b2}
\end{eqnarray}
Combining Eq.~\ref{sum_ai}, \ref{a3}, \ref{b3}
and~\ref{a1_minus_a2=b1_minus_b2},
\begin{eqnarray}
  &a_2 = N-a_1-n^{(A)},\\
  &b_1 = a_1 + \frac12\left(n^{(A)}-n^{(B)}\right),\\
  &b_2 = N-a_1-\frac12\left(n^{(A)}+n^{(B)}\right),
\end{eqnarray}
so that, by plugging into Eq.~\ref{a1a2a3} and~\ref{b1b2b3}, we obtain
\begin{eqnarray}
  \frac{\Pr(A\to B)}{\Pr(B\to A)} =
  \left(2\sqrt{p_+p_-}\right)^{n^{(B)}-n^{(A)}}.
  \label{lhs_proof}
\end{eqnarray}
Comparing Eq.~\ref{voter_beta}, \ref{rhs_proof} and~\ref{lhs_proof}
proves Eq.~\ref{det_bal}.

\section{The voter model with random asynchronous updates}
\label{asynch_section}
Not only the voter model with perfectly synchronous updates of
opinions can be mapped to an even or odd Ising model.
We will now argue that, by defining the spins as in
Eq.~\ref{voter_sigma}, we can also interpret asynchronous updates of
randomly selected single opinions in terms of an Ising Hamiltonian.
While the synchronous case, as shown in the previous section,
corresponds to a positive spin interaction $J$ and zero magnetic field
$H$, the asynchronous case leads to $J=0$ and, in general, $H\neq0$
for the following reason.

\begin{figure}
  \begin{center}
    \includegraphics[width=7cm]{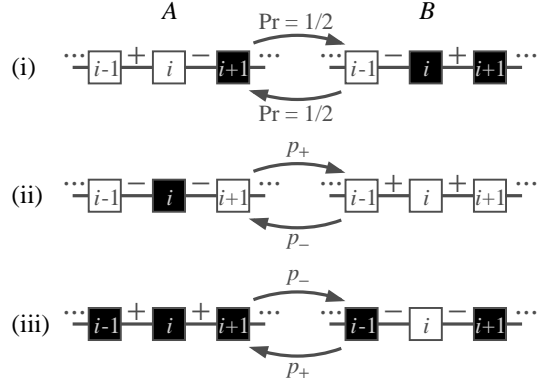}
    \caption{
      Transition probabilities for asynchronous updates.
      Depicted are three representative cases where only opinion $i$
      changes between states $A$ and $B$.
      All other cases can be generated by inverting all opinions (from
      white to black and vice versa) and/or interchanging the order of
      the chain so that $i-1$ and $i+1$ trade places.
    }
    \label{asynch}
  \end{center}
\end{figure}

Suppose opinion $\omega_i$ is chosen to be updated. The
probability to have opinion $\Omega$ in the next time step is given by
Eq.~\ref{voter_model_Pr} while all other opinions remain unchanged.
Then the only spins affected are $\sigma_{i-1}$ and $\sigma_i$ so that
we can ignore the rest of the chain.
If $\omega_i$ changes between states $A$ and $B$, then we can
distinguish the three cases depicted in Fig.~\ref{asynch}: either
\begin{enumerate}
\item {$\Pr(A\to B)=\Pr(B\to A)=\frac12$ 
    and
    $\sigma_i^{(A)}+\sigma_{i+1}^{(A)}=\sigma_i^{(B)}+\sigma_{i+1}^{(B)}=0$
    or }
\item {$\Pr(A\to B)=p_+$ and $\Pr(B\to A)=p_-$ and
    $\sigma_i^{(A)}+\sigma_{i+1}^{(A)} = -\sigma_i^{(B)}-\sigma_{i+1}^{(B)}
    = -2$ or}
\item {$\Pr(A\to B)=p_-$ and $\Pr(B\to A)=p_+$ and
    $\sigma_i^{(A)}+\sigma_{i+1}^{(A)} =
    -\sigma_i^{(B)}-\sigma_{i+1}^{(B)} = 2$.}
\end{enumerate}
In summary, we can write all of these cases as 
\begin{align}
  \frac{\Pr(A\to B)}{\Pr(B\to A)} =
  \left(\frac{p_+}{p_-}\right) ^
  {-\frac14\left(\sum\sigma_i^{(A)}-\sum\sigma_i^{(B)}\right)},
\end{align}
which is of the form of Eq.~\ref{det_bal} with the energy
$E=-H\sum_i\sigma_i$ and inverse temperature
\begin{align}
  \beta = \frac{\ln(p_+/p_-)}{4H}.
  \label{betaH}
\end{align}

If $p_+\in(\frac12,1)$, we must have $H>0$ to obtain a positive
temperature. Asynchronous opinion updates then tend to favour the
states depicted in Fig.~\ref{voter_stat}(a) and (b) where the spins
are all positive.
Generally, the states in Fig.~\ref{voter_stat}(c) are suppressed when
$p_+>1/2$ and the dynamic rule mixes synchronous and asynchronous
updates (e.g.\ by updating a fraction of the opinions in every update
as in Ref.~\cite{SznajdKrupa06}).
The opposite is true for $p_+<1/2$ where asynchronous updates generate
sequences of alternating opinions and suppress unanimity.
All cases, however, have in common that the periodic boundary
conditions in the opinions generate an even number of negative spins,
resulting in an even (odd) Ising model for even (odd) $N$.

Whether synchronous, asynchronous or partially synchronous updates are
more realistic depends on the situation one wishes to model.
Asynchronous updates have a long tradition in physics (e.g.\ the
Glauber or Metropolis rules for dynamic Ising models), but synchronous
updates, especially in the context of stochastic cellular
automata~\cite{Wolfram83}, have also been investigated (for example
in~\cite{Nowak_etal94, Fernandez_etal11, Varga_etal14}).
If agents can only make decisions at discrete times (e.g.\ only at the
end of a business day or if biological populations exhibit strongly
peaked cyclic activity~\cite{GodfrayHassell89}), then synchronous
or partially synchronous updates are more applicable.
Here we do not intend to argue for any particular update rule. 
Generally, one has to be humble about social or economic
interpretations of such simple rules~\cite{Gallegati_etal06} because
true opinion dynamics is far more complex.
Our focus here is rather on the model's structural properties
in order to motivate how the even and odd Ising models can arise
from another two-state model.

\section{Partition function}
\label{partition_function}

We denote the partition function for the even and odd Ising model
by $Z_e$ and $Z_o$ respectively,
\begin{eqnarray}
  &Z_e = \sum_{\substack{
      \boldsymbol{\sigma}\text{ with}\atop \text{even
      }N_+(\boldsymbol{\sigma})}
  }
  e^{-\beta 
    E(\boldsymbol{\sigma})},
  \label{Z_e_def}\\
  &Z_o = \sum_{\substack{\boldsymbol{\sigma}\text{
        with}\atop \text{odd }N_+(\boldsymbol{\sigma})}}e^{-\beta
    E(\boldsymbol{\sigma})}.
  \label{Z_o_def}
\end{eqnarray}
If we associate a spin $\sigma_i=1$ with the bra vector $\Bra{+1} =
(1,0)$ and $\sigma_i=-1$ with $\Bra{-1} = (0,1)$, we can write $Z_e$
with the transfer matrix of the unconstrained
model~\cite{KramersWannier41}
\begin{eqnarray}
  \mathbf{P} = 
  \left( {\begin{array}{cc}
        e^{\beta(J+H)} & e^{-\beta J}\\
        e^{-\beta J} & e^{\beta(J-H)}\\
      \end{array} } \right)
  \label{P}\
\end{eqnarray}
as
\begin{eqnarray}
\fl
  Z_e = \sum_{\substack{\boldsymbol{\sigma}\text{ with}\atop\text{even }N_+(\boldsymbol{\sigma})}} \Bra{\sigma_1}
 \mathbf{P} \Ket{\sigma_2} \Bra{\sigma_2}
 \mathbf{P} \Ket{\sigma_3} \ldots \Bra{\sigma_N} \mathbf{P}
 \Ket{\sigma_1}
    = \Tr\left(\sum_{\substack{\boldsymbol{\sigma}\text{
         with}\atop\text{even }N_+(\boldsymbol{\sigma})}}
   \Ket{\sigma_1}\Bra{\sigma_1} \mathbf{P} \ldots
   \Ket{\sigma_N}\Bra{\sigma_N} \mathbf{P}
\right).
\end{eqnarray}
Similarly,
\begin{eqnarray}
  Z_o = \Tr\left(\sum_{\substack{\boldsymbol{\sigma}\text{
          with}\atop\text{odd }N_+(\boldsymbol{\sigma})}}
   \Ket{\sigma_1}\Bra{\sigma_1} \mathbf{P} \ldots
   \Ket{\sigma_N}\Bra{\sigma_N} \mathbf{P}
\right).
\end{eqnarray}
Let us define
\begin{eqnarray}
  \mathbf{M}_e = \sum_{\substack{\boldsymbol{\sigma}\text{
        with}\atop\text{even }N_+(\boldsymbol{\sigma})}}
  \Ket{\sigma_1}\Bra{\sigma_1} \mathbf{P} \ldots
  \Ket{\sigma_N}\Bra{\sigma_N} \mathbf{P},\\
  \mathbf{M}_o = \sum_{\substack{\boldsymbol{\sigma}\text{
        with}\atop\text{odd }N_+(\boldsymbol{\sigma})}}
  \Ket{\sigma_1}\Bra{\sigma_1} \mathbf{P} \ldots
  \Ket{\sigma_N}\Bra{\sigma_N} \mathbf{P}.
\end{eqnarray}
Induction on $N$ proves
\begin{eqnarray}
  \left( {\begin{array}{cc}
        \mathbf{M}_e, & \mathbf{M}_o\\
        \mathbf{M}_o, & \mathbf{M}_e\\
      \end{array} } \right) =
  \left( {\begin{array}{cc}
        \Ket{-1}\Bra{-1}\mathbf{P}, & 
        \Ket{+1}\Bra{+1}\mathbf{P} \\
        \Ket{+1}\Bra{+1}\mathbf{P}, & 
        \Ket{-1}\Bra{-1}\mathbf{P}\\
      \end{array} } \right)^N.
\end{eqnarray}
With the definition
\begin{eqnarray}
  \fl
  \mathbf{Q} = \left( {\begin{array}{cc}
        \Ket{-1}\Bra{-1}\mathbf{P}, & 
        \Ket{+1}\Bra{+1}\mathbf{P} \\
        \Ket{+1}\Bra{+1}\mathbf{P}, & 
        \Ket{-1}\Bra{-1}\mathbf{P}\\
      \end{array} } \right) =
  \left( {\begin{array}{cccc}
        0 & 0 & e^{\beta(J+H)} & e^{-\beta J}\\
        e^{-\beta J} & e^{\beta(J-H)} & 0 & 0\\
        e^{\beta(J+H)} & e^{-\beta J} & 0 & 0\\
        0 & 0 & e^{-\beta J} & e^{\beta(J-H)}\\
        \end{array} } \right)
  \label{Q}\\
\end{eqnarray}
we can write
\begin{eqnarray}
  Z_e = \Tr\left(\mathbf{M}_e\right) =
  \frac12\Tr\left(\mathbf{Q}^N\right).
  \label{Ze_as_Tr}
\end{eqnarray}
To simplify the notation further, we introduce
\begin{eqnarray}
  &x = \beta H,\\
  &y = \beta J.
\end{eqnarray}
The eigenvalues of $\mathbf{Q}$ are then
\begin{eqnarray}
  &\lambda_{1,2} = e^y\left(\cosh x \pm \sqrt{\sinh^2x +
      e^{-4y}}\right),
  \label{lambda_12}\\
  &\lambda_{3,4} = e^y\left(-\sinh x \pm \sqrt{\cosh^2x -
      e^{-4y}}\right).
  \label{lambda_34}
\end{eqnarray}
Consequently,
\begin{eqnarray}
  Z_e = \frac12\left(\lambda_1^N + \lambda_2^N + \lambda_3^N +
    \lambda_4^N\right).
  \label{Z_e}
\end{eqnarray}
We can derive $Z_o$ as follows.
The eigenvalues $\lambda_{1,2}$ are also the eigenvalues of
$\mathbf{P}$ and therefore the partition function of the unconstrained
Ising model is $Z_u = \lambda_1^N+\lambda_2^N$.
Moreover $Z_u = Z_e+Z_o$ so that
\begin{eqnarray}
  Z_o = \frac12\left(\lambda_1^N + \lambda_2^N - \lambda_3^N -
    \lambda_4^N\right).
  \label{Z_o}
\end{eqnarray}
Because $\lambda_1$ is the leading eigenvalue, we find in the
thermodynamic limit (i.e.\ $N\to\infty$) with fixed $x$ and $y$ that $Z_e
\propto Z_o \propto Z_u \propto \lambda_1^N$.
As a consequence, all equilibrium properties of the even and odd Ising
models converge to the same limits as the unconstrained model.
However, we will analytically derive in
Sec.~\ref{leading_order_correction} different scaling limits for
$N\to\infty$ when temperature and magnetic field go to their critical
value (i.e.\ zero) such that $Ne^{-2y}$ and $N\sinh x$ are asymptotically
constants.
For this purpose, it will be instructive to derive first some exact
formulae for finite $N$.

\section{Magnetization and Susceptibility}
\label{magn_susc}

We first calculate the mean magnetization per spin 
\begin{eqnarray}
  \langle m\rangle \equiv \frac{\langle\sum_i\sigma_i\rangle}N =
  \frac1N\frac{\partial}{\partial x}\ln Z,
  \label{m_from_Z}
\end{eqnarray}
where $Z$ is the partition function of the model in question and the
angle brackets denote the ensemble average.
With the auxiliary functions
\begin{eqnarray}
  &s_1(x,y) = \frac{\sinh x}{\sqrt{\sinh^2x+e^{-4y}}},
  \label{s1}\\
  &c_1(x,y) = \frac{\cosh x}{\sqrt{\cosh^2x-e^{-4y}}},
  \label{c1}
\end{eqnarray}
we can write Eq.~\ref{m_from_Z} as
\begin{eqnarray}
  \langle m\rangle_{e,o}(x,y) =
  \frac{s_1\left(\lambda_1^N-\lambda_2^N\right) \mp
    c_1\left(\lambda_3^N-\lambda_4^N\right)}
  {\lambda_1^N+\lambda_2^N\pm\lambda_3^N\pm\lambda_4^N},
  \label{m_eo}
\end{eqnarray}
where the upper signs apply to the even and the lower signs to the odd
model.

\begin{figure}
  \begin{center}
    \includegraphics[width=15.5cm]{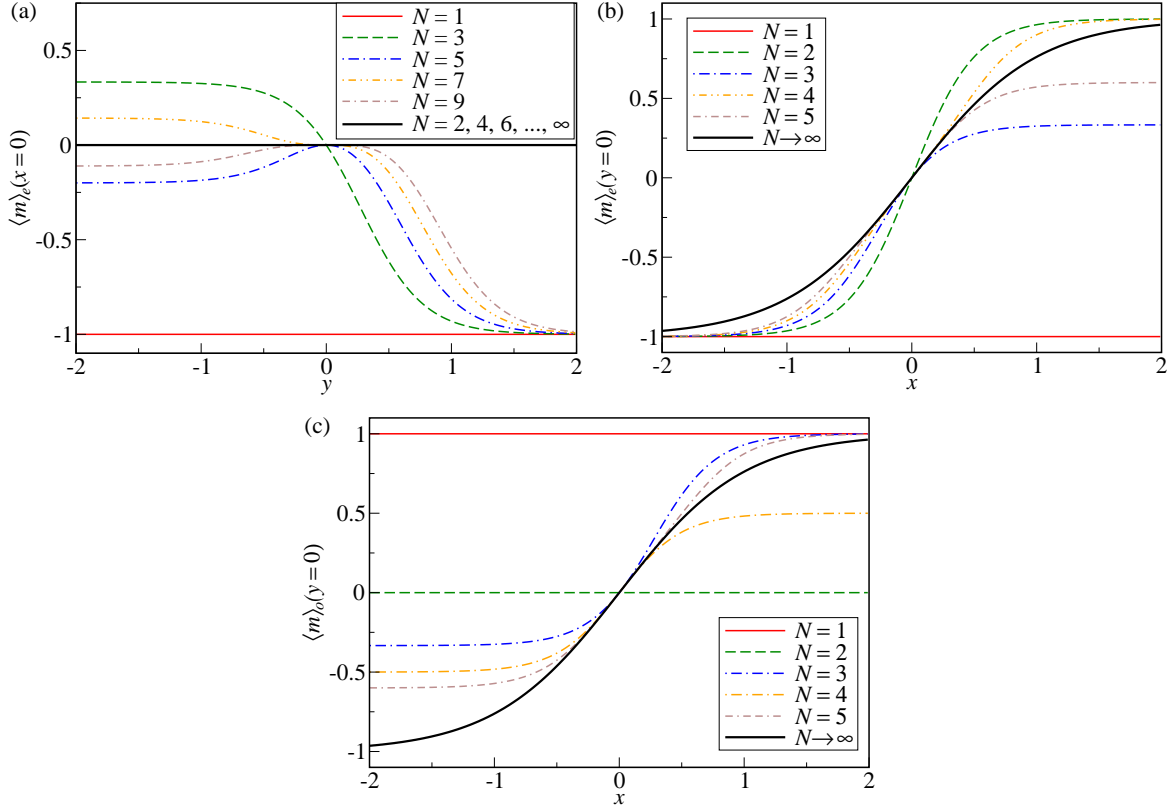}
    \caption{
      The mean magnetization $\langle m_e\rangle$ (a) as a function of
      $y$ when $x=0$, (b) as a function of $x$ when $y=0$. (c) The mean
      magnetization $\langle m_o\rangle$ for $y=0$.
    }
    \label{magn}
  \end{center}
\end{figure}

In the special case $x=0$, applicable to the synchronous voter model,
we insert the eigenvalues from Eq.~\ref{lambda_12} and~\ref{lambda_34}
(Fig.~\ref{magn}a)
\begin{eqnarray}
  \langle m\rangle_{e,o}(x=0,y) =
  \begin{cases}
    \langle m\rangle_u(x=0,y) = 0 & \text{if $N$ is even},\\
    \mp\frac{e^y}{\cosh^Ny+\sinh^Ny}
    \left(\frac{\sinh(2y)}2\right)^{(N-1)/2}
    & \text{if $N$ is odd.}
  \end{cases}
 \label{meo_x0}
\end{eqnarray}
Hence, for odd $N$, even when there is no external magnetic field
(i.e.\ $H=0$), the magnetization is generally different from zero.
This phenomenon can be intuitively explained.
The constraint of an even number of positive spins prevents for
odd $N$ a ground state with perfectly aligned
positive spins.
However, the state with $\sigma_1=\ldots=\sigma_N=-1$ is permitted and
therefore the mean magnetization in the limit $y\to\infty$ is $-1$.
The same argument applies with opposite signs to the odd model.
In the antiferromagnetic limit (i.e.\ $y\to-\infty$) the neighbouring
spins prefer to be in opposite directions, but an odd $N$ forces at
least one pair to point in the same direction and thus $m_{e,o} =
(-1)^{(N\pm1)/2}/N$.

The relevant case for the asynchronous voter model is $y=0$ where the
interactions between spins are negligible compared to the external
magnetic field,
\begin{eqnarray}
  \langle m\rangle_{e,o}(x,y=0) = 
  \frac{\sinh(2x)}2 \times \frac{\cosh^{N-2}x\pm(-1)^N\sinh^{N-2}x}
  {\cosh^Nx\pm(-1)^N\sinh^Nx}.\label{meo_y0}
\end{eqnarray}
The functions are plotted in Fig.~\ref{magn}(b) and (c).
In the unconstrained model the magnetization $\langle
m\rangle_u(x,y=0) = \tanh x$ is independent of $N$.
However, in the even and odd models, the constraints on the number of
spins acts as an effective interaction so that the partition function
does not factorize although $J=0$.
Consequently, the magnetization of Eq.~\ref{meo_y0} depends on $N$. 

The fluctuations in the magnetization are measured by the
susceptibility per spin
\begin{eqnarray}
  \chi \equiv \beta N\left(\langle m^2\rangle - \langle
    m\rangle^2\right) = \frac{\partial\langle m\rangle}{\partial H}.
\end{eqnarray}
Taking the derivative of Eq.~\ref{m_eo} for general $x$ and $y$ is in
principle possible, but leads to rather lengthy expressions.
We focus here instead directly on the two special cases $x=0$ and
$y=0$.

\begin{figure*}
  \begin{center}
    \includegraphics[width=15.5cm]{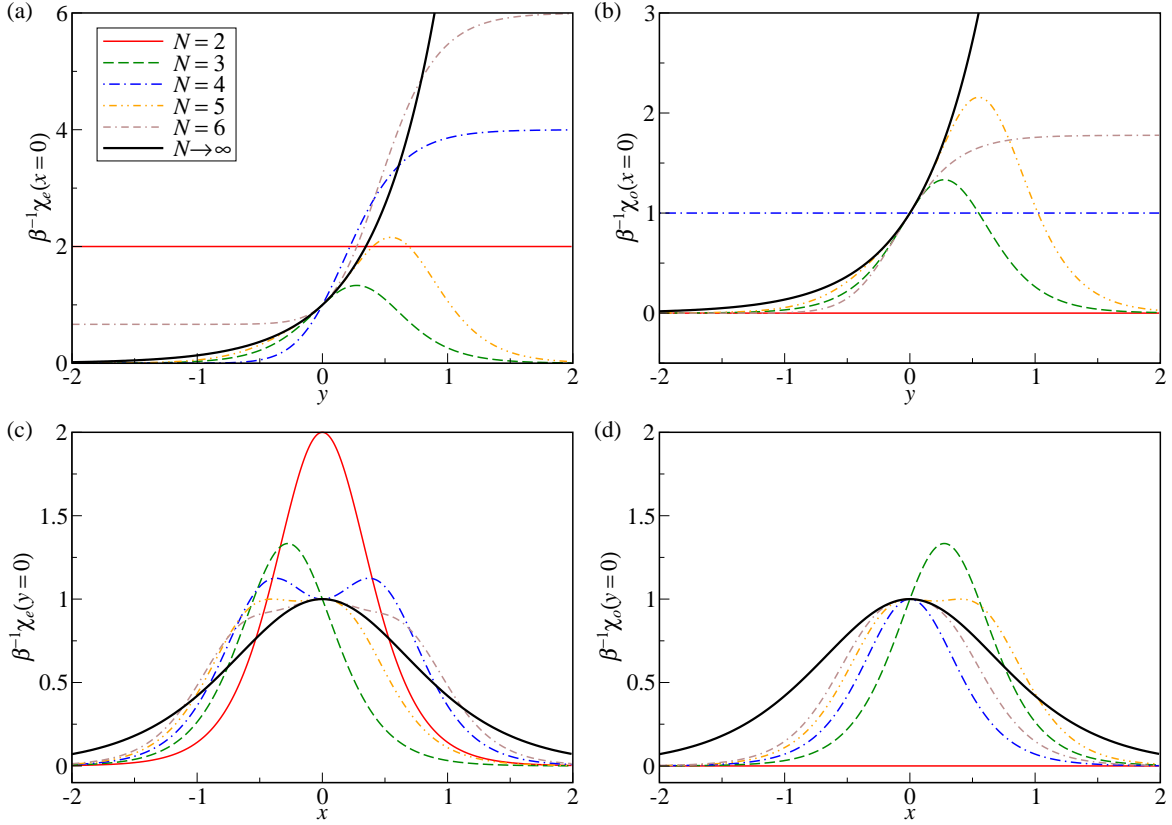}
    \caption{
      The susceptibility $\chi$ divided by the inverse temperature
      $\beta$ for (a), (b) zero magnetic field $H$ as a function of
      $y=\beta J$ and (c), (d) zero spin interaction $J$ as a function
      of $x=\beta H$.
      Panels (a) and (c) show the results for the even Ising model,
      (b) and (d) for the odd model.
    }
    \label{susc}
  \end{center}
\end{figure*}

For $x=0$ (i.e.\ in the absence of an external magnetic field),
\begin{eqnarray}
  \chi_{e,o}(x=0,y) = 
  \begin{cases}
    \frac{\beta e^{2y}\left(\cosh^Ny-\sinh^Ny\pm2^{-N/2}N\sinh^{N/2-1}(2y)
      \right)}{\left(\cosh^{N/2}y\pm\sinh^{N/2}y\right)^2} & \text{if
      $N$ is even,}\\
    \frac{\beta
      e^{2y}\left(\cosh^{2N}y-\sinh^{2N}y-2^{1-N}N\sinh^{N-1}(2y)\right)}
    {\left(\cosh^Ny+\sinh^Ny\right)^2}
      & \text{if $N$ is odd,}
    \end{cases}
\label{chieo_x0}
\end{eqnarray}
compared to $\chi_u=\beta
e^{2y}(\cosh^Ny-\sinh^Ny)/(\cosh^Ny+\sinh^Ny)$.
Plotting $\chi_{e,o}$ in Fig.~\ref{susc}a and
\ref{susc}b, the most striking feature for odd $N$ is
$\lim_{y\to\infty}\chi_{e,o}=0$, whereas the unconstrained Ising model
(and the even model for even $N$) reaches in this limit its maximum
susceptibility $\beta N$.
The reason is that, as already mentioned, the even and odd models for
odd $N$ only have one ground state each, but the unconstrained model
has two.

For the odd model with even $N$ we observe yet another interesting
phenomenon.
The susceptibility reaches its maximum in the limit $y\to\infty$, but
with a smaller value than the unconstrained or even models, namely
$\lim_{y\to\infty}\chi_o = \beta(N^2-4)/(3N)$.
The explanation is that the perfectly aligned ground states of the
unconstrained models are not permitted.
Therefore, the states of minimum energy in the odd model are the first
excited states of the unconstrained model whose magnetization is not
confined to the extreme values $m=\pm1$.

In the case of no internal interactions (i.e.\ $y=0$),
\begin{eqnarray}
  \fl\chi_{e,o}(x,y=0) =  
\beta\left(
    \frac{\cosh^{N-2}x \mp (-1)^N\sinh^{N-2}x} {\cosh^Nx \pm
      (-1)^N\sinh^Nx}
    \pm 
    \frac{(-1)^NN\sinh^{N-2}(2x)} {2^{N-2}\left(\cosh^Nx \pm
        (-1)^N\sinh^Nx\right)^2}\right),
   \label{chieo_y0}
\end{eqnarray}
while $\chi_u=\beta\cosh^{-2}x$.
We plot $\chi_{e,o}$ in Fig.~\ref{susc}c and~\ref{susc}d.
For odd $N$, they satisfy $\chi_e(x,0)=\chi_o(-x,0)$ because in this
case the odd model is equivalent to the even model with flipped signs
of spins and magnetic field.
If $N$ is even, $\chi_e$ and $\chi_o$ are intrinsically symmetric, but
with larger values in the tails of the even model because
$\chi_e/\chi_o \to 3N/(N-2)$ as $|x|\to\infty$ and $y=0$.

\section{Nearest-neighbour correlations, internal energy and heat
  capacity}
\label{corr_inten_heatcap}
If we replace in Eq.~\ref{m_from_Z} the partial derivative with
respect to $x$ by differentiation with respect to $y$, we obtain the
mean nearest-neighbour correlation
\begin{eqnarray}
  \langle g_1\rangle \equiv
  \frac1N\left\langle\sum_i\sigma_i\sigma_{i+1}\right\rangle =
  \frac1N\frac{\partial}{\partial y}\ln Z.
  \label{g1_from_Z}
\end{eqnarray}
If we define the functions
\begin{eqnarray}
  s_2(x,y) = 2e^{-3y}\left(\sinh^2x+e^{-4y}\right)^{-1/2},\\
  c_2(x,y) = 2e^{-3y}\left(\cosh^2x-e^{-4y}\right)^{-1/2},
\end{eqnarray}
then
\begin{eqnarray}
  \langle g_1\rangle_{e,o}(x,y) =
  1 +
  \frac{s_2\left(\lambda_2^{N-1}-\lambda_1^{N-1}\right) \pm
    c_2\left(\lambda_3^{N-1}-\lambda_4^{N-1}\right)}
  {\lambda_1^N + \lambda_2^N \pm \lambda_3^N \pm
      \lambda_4^N}.
\end{eqnarray}

Without external magnetic field,
\begin{eqnarray}
  \langle g_1\rangle&_{e,o}(x=0,y) =
  &\begin{cases}
    1 - \frac{\cosh^{N/2-1}y \mp \sinh^{N/2-1}y}
    {e^y\left(\cosh^{N/2}y \pm \sinh^{N/2}y\right)} & \text{if $N$ is
        even}, \\
  \langle g_1\rangle_u(x=0,y) = 1 + \frac{\sinh^{N-1}y - \cosh^{N-1}y} {e^y\left(\cosh^Ny +
          \sinh^Ny\right)} & \text{if $N$ is odd}.
  \end{cases}
  \label{g1eo_x0}
\end{eqnarray}
If $N$ is odd, $\langle g_1\rangle_{e,o}(x=0,y)$ is equal to the
correlation in the unconstrained model for the following reason.
The spin configurations in the unconstrained model can be
divided into two sets: one set containing all configurations of the
even model and another set with all odd-numbered states.
We can map every element in one set uniquely to the configuration in
the other set that has all spins inverted. 
Because the sum of Eq.~\ref{g1_from_Z} is invariant if all spins are
simultaneously flipped, the average correlations must be equal in both
sets.
The same argument cannot be applied to even $N$, however, because
inverting the spins in the even or odd set generates another spin in
the same set.
As a consequence, $\langle g_1\rangle_{e}$ and $\langle g_1\rangle_{o}$
are in this case different functions.

With a magnetic field, but with vanishing spin interactions,
\begin{eqnarray}
  \langle g_1\rangle_{e,o}(x, y=0) = 
  1 - \frac{\cosh^{N-2}x \mp (-1)^N\sinh^{N-2}x} {\cosh^Nx \pm
    (-1)^N\sinh^Nx},
  \label{g1eo_y0}
\end{eqnarray}
compared to $\langle g_1\rangle_u=\tanh^2x$.
For odd $N$, we find $\langle g_1\rangle_e(x,0) =
\langle g_1\rangle_o(-x,0)$ for the same reason as discussed after the
corresponding Eq.~\ref{chieo_y0} for the susceptibility.
We also find again that, for even $N$, $\langle g_1\rangle_e(x,0) =
\langle g_1\rangle_e(-x,0)$ and
$\langle g_1\rangle_o(x,0) = \langle g_1\rangle_o(-x,0)$.
Expanding Eq.~\ref{g1eo_y0}, however, shows that the limits for
$|x|\to\infty$ and even $N$ are different: $\langle g_1\rangle_e\to1$,
but $\langle g_1\rangle_o\to1-4/N$.
The intuition behind this result is that a strong magnetic field can
perfectly align the spins in the even, but not in the odd model.

Closely related to the nearest-neighbour correlations is the internal
energy (i.e.\ ensemble average of the Hamiltonian) per spin
\begin{eqnarray}
  U \equiv \frac{\langle E\rangle}{N} =
  -\frac1N\frac{\partial}{\partial\beta}\ln Z.
\end{eqnarray}
In general, the calculation yields rather lengthy expressions. 
However, if $H=0$, then $U_{e,o} = -J\langle g_1\rangle_{e,o}(x=0,y)$.
If, on the other hand, $J$ vanishes, then $U_{e,o} = -H
\langle m\rangle_{e,o}(x,y=0)$.
Using our earlier results of Eq.~\ref{meo_y0} and~\ref{g1eo_x0},
\begin{eqnarray}
   U_{e,o}(x=0,y) =
  \begin{cases}
    J\left(\frac{\cosh^{N/2-1}y \mp \sinh^{N/2-1}y}
    {e^y\left(\cosh^{N/2}y \pm \sinh^{N/2}y\right)} -1\right) & \text{if $N$ is
        even}, \\
      U_u(x=0,y) = J\left(\frac{\cosh^{N-1}y - \sinh^{N-1}y} {e^y\left(\cosh^Ny +
          \sinh^Ny\right)} -1\right) & \text{if $N$ is odd},
  \end{cases}
  \label{Ueo_x0}
\end{eqnarray}
\begin{eqnarray}
  U&_{e,o}(x,y=0) = 
  &-\frac{H\sinh(2x)}2 \times \frac{\cosh^{N-2}x\pm(-1)^N\sinh^{N-2}x}
  {\cosh^Nx\pm(-1)^N\sinh^Nx}\label{Ueo_y0},
\end{eqnarray} 
while $U_u(x,y=0)=-H\tanh x$.

Taking another derivative of $\ln Z$ with respect to $\beta$ gives us
the heat capacity per spin, which measures the fluctuations in the
energy,
\begin{eqnarray}
  C \equiv \frac{k_B\beta^2}N\left(\langle E^2\rangle - \langle
    E\rangle^2\right) = -k_B\beta^2\frac{\partial U}{\partial\beta}.
\end{eqnarray}
For vanishing $H$, we can use $(\partial U)/(\partial\beta) =
J(\partial U)/(\partial y)$ and Eq.~\ref{Ueo_x0}.
If $J=0$, then $C=k_B\beta H^2\chi$, so the heat capacity follows
directly from Eq.~\ref{chieo_y0},
\begin{eqnarray}
  \fl C_{e,o}(x=0,y) = \label{Ceo_x0}\\
  \fl \hspace{0.5cm}\begin{cases}
    \frac{k_B\beta^2J^2}{\cosh^{N/2}y \pm \sinh^{N/2}y}\left(
      \cosh^{N/2-2}y \mp \sinh^{N/2-2}y \pm \frac {N\sinh^{N/2-2}(2y)}
      {2^{N/2-1}\left(\cosh^{N/2}y \pm \sinh^{N/2}y\right)}
    \right) & \text{if $N$ is even,}\\
    C_u(x=0,y) =\frac{k_B\beta^2J^2}{\cosh^Ny + \sinh^Ny}\left(
      \cosh^{N-2}y - \sinh^{N-2}y + \frac{N\sinh^{N-2}(2y)}
      {2^{N-2}\left(\cosh^Ny + \sinh^Ny\right)}
    \right) & \text{if $N$ is odd,}
  \end{cases}\nonumber
\end{eqnarray}
\begin{eqnarray}
  \fl C_{e,o}(x,y=0) =\label{Ceo_y0}\\
  \fl \hspace{0.5cm} k_B\beta^2H^2\left(
    \frac{\cosh^{N-2}x \mp (-1)^N\sinh^{N-2}x} {\cosh^Nx \pm
      (-1)^N\sinh^Nx}
    \pm 
    \frac{(-1)^NN\sinh^{N-2}(2x)} {2^{N-2}\left(\cosh^Nx \pm
        (-1)^N\sinh^Nx\right)^2}\right),
  \nonumber
\end{eqnarray}
approaching $C_u(x=0,y)=k_B\beta^2H^2\cosh^{-2}(x)$ in the thermodynamic
limit.

\section{Correlation function}
\label{corrfunc_section}
We can generalize the calculation in the previous section to find the
correlation between $k$-th nearest neighbours.
For this purpose we make the spin interactions $J$ in an auxiliary
Hamiltonian $\tilde{E}$ dependent on the position $i$, but for
simplicity's sake we drop the magnetic field,
\begin{eqnarray}
  \tilde{E}(\boldsymbol{\sigma}) =
  -\sum_{i=1}^N{J_i\sigma_i\sigma_{i+1}}.
\end{eqnarray}
Applying the same line of reasoning that led us to Eq.~\ref{Ze_as_Tr}, we
can show that the partition function for the Hamiltonian $\tilde{E}$
in the case of even $N^+$ is
\begin{eqnarray}
  \tilde{Z}_e =
  \frac12\Tr\left(\prod_{i=1}^N\mathbf{Q}_i\right),
  \label{tildeZ_as_Tr}
\end{eqnarray}
where
\begin{eqnarray}
  \mathbf{Q}_i = 
  &\left( {\begin{array}{cccc}
        0 & 0 & e^{\beta J_i} & e^{-\beta J_i}\\
        e^{-\beta J_i} & e^{\beta J_i} & 0 & 0\\
        e^{\beta J_i} & e^{-\beta J_i} & 0 & 0\\
        0 & 0 & e^{-\beta J_i} & e^{\beta J_i}\\
        \end{array} } \right)
\end{eqnarray}
plays the role of the transfer matrix of Eq.~\ref{Q}.
The matrices $\mathbf{Q}_i$ do not commute and therefore we cannot
simultaneously diagonalize them for computing the trace in
Eq.~\ref{tildeZ_as_Tr}.
However, the product $\mathbf{Q}_i\mathbf{Q}_{i+1}$ commutes with
$\mathbf{Q}_{i+2}\mathbf{Q}_{i+3}$. 
These products are diagonalized as $\mathbf{R}\mathbf{Q}_i\mathbf{Q}_{i+1}\mathbf{R}$
by the matrix of eigenvectors
\begin{eqnarray}
  \mathbf{R} =
  \frac12\left( {\begin{array}{rrrr}
                   1 & 1 & 1 & 1\\
                   1 & -1 & 1 & -1\\
                   1 & 1 & -1 & -1\\
                   1 & -1 & -1 & 1\\
                 \end{array} } \right) 
  = \mathbf{R}^{-1},
\end{eqnarray}
and the corresponding eigenvalues are
\begin{eqnarray}
  &\left(\mathbf{R}\mathbf{Q}_i\mathbf{Q}_{i+1}\mathbf{R}\right)_{11}
    = 4\cosh(\beta J_i)\cosh(\beta J_{i+1}),\\
  &\left(\mathbf{R}\mathbf{Q}_i\mathbf{Q}_{i+1}\mathbf{R}\right)_{22}
    = 4\sinh(\beta J_i)\sinh(\beta J_{i+1}),\\
  &\left(\mathbf{R}\mathbf{Q}_i\mathbf{Q}_{i+1}\mathbf{R}\right)_{33}
    = 4\sinh(\beta J_i)\cosh(\beta J_{i+1}),\\
  &\left(\mathbf{R}\mathbf{Q}_i\mathbf{Q}_{i+1}\mathbf{R}\right)_{44}
    = 4\cosh(\beta J_i)\sinh(\beta J_{i+1}).
\end{eqnarray}
If $N$ is even, it follows that
\begin{eqnarray}
  \fl \tilde{Z}_{e,\text{ even $N$}} = 2^{N-1}\bigg(
    \prod_{i=1}^{N}\cosh(\beta J_i) + \prod_{i=1}^{N}\sinh(\beta
    J_i) + \\
  \hspace{0.5cm}\prod_{i=1}^{N/2}\cosh(\beta J_{2i-1})\sinh(\beta J_{2i}) +
    \prod_{i=1}^{N/2}\sinh(\beta J_{2i-1})\cosh(\beta J_{2i})
    \bigg),\nonumber
\end{eqnarray}
while for odd $N$ the partition function is half of the unconstrained
model's partition function
\begin{eqnarray}
  \tilde{Z}_{e,\text{ odd $N$}} = \frac12\tilde{Z}_u = 2^{N-1}\left(
  \prod_{i=1}^{N}\cosh(\beta J_i) + \prod_{i=1}^{N}\sinh(\beta J_i)
  \right).
\end{eqnarray}
For the odd model, we can apply $\tilde{Z}_o = \tilde{Z}_u-\tilde{Z_e}$.
The disconnected correlation function $\langle g_k\rangle$ can now be
computed as
\begin{eqnarray}
  \langle g_k\rangle = 
   \frac1N\left\langle\sum_i\sigma_i\sigma_{i+k}\right\rangle = \left[\frac1{\beta^k\tilde{Z}} 
    \frac{\partial}{\partial J_1} \frac{\partial}{\partial J_2}
    \ldots \frac{\partial}{\partial J_k}
    \tilde{Z}\right]_{J_1=\ldots=J_N=J}
\end{eqnarray}
with the final result
\begin{eqnarray}
  \fl \langle g_k\rangle_{e,o} =
  \begin{cases}
  \frac{
  \cosh^{N-k}y\sinh^ky +
  \sinh^{N-k}y\cosh^ky \pm
  2^{1-N/2}\cosh(2y)\sinh^{N/2-1}(2y)
  }
  {\left(\cosh^{N/2}y\pm\sinh^{N/2}y\right)^2}&\text{if $N$ even, $k$ odd,}\\
  \frac{
  \cosh^{N-k} y\sinh^k y +
  \sinh^{N-k} y\cosh^k y \pm
  2^{1-N/2}\sinh^{N/2}(2y)
  }
  {\left(\cosh^{N/2}y\pm\sinh^{N/2}y\right)^2}&\text{if $N$ even, $k$ even,}\\
  \langle g_k\rangle_u = \frac{\cosh^{N-k}y\,\sinh^ky +
  \sinh^{N-k}y\,\cosh^ky} {\cosh^Ny+\sinh^Ny}&\text{if
  $N$ odd}. 
  \end{cases}
\rule[-1.5em]{10pt}{0pt}
\end{eqnarray}
Taking the limit $N\to\infty$ while keeping $y$ and $k$ fixed, all
three cases have the same asymptotic value $\lim_{N\to\infty} \langle
g_k\rangle_{e,o,u} = \tanh^ky$ and the correlation length is hence
$\xi=-[\ln(\tanh y)]$.
The divergence at $y=0$ can be expressed as a power law in the reduced
temperature~\cite{NelsonFisher75}
\begin{eqnarray}
  T_r = e^{-2y}
\end{eqnarray}
because near $T_r=0$
\begin{eqnarray}
  \xi \approx {2T_r}^{-1} \propto T_r^{-\nu},
  \label{crit_exp_nu}
\end{eqnarray}
where the critical correlation length exponent satisfies $\nu=1$ in
the unconstrained, even and odd model.

\section{Approach to the thermodynamic limit}
\label{leading_order_correction}

\begin{figure}
  \begin{center}
    \includegraphics[width=7.75cm]{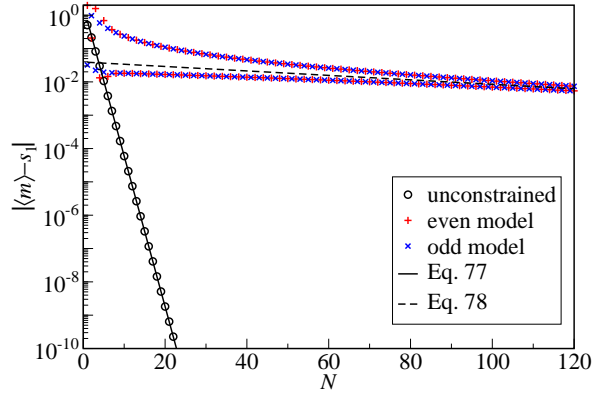}
    \caption{
      The difference between the magnetization $\langle m\rangle$ and
      $s_1$ for a finite chain of length $N$, $x=0.5$ and $y=1$.
      We note that $s_1$, defined in Eq.~\ref{s1}, is the thermodynamic
      limit of $\langle m\rangle$.
      Circles, $+$ and $\times$ symbols are exact results. The solid
      and dashed lines are the leading-order approximations of
      Eq.~\ref{mu_lead_ord} and \ref{meo_lead_ord}.
      The constrained models converge much more slowly than their
      unconstrained counterpart.
    }
    \label{lead_ord_corr}
  \end{center}
\end{figure}

It is not surprising that $\nu$ does not depend on whether we constrain
the number of positive spins to even or odd values or have no such
constraint.
We have already pointed out the reason after Eq.~\ref{Z_o}: the
thermodynamic limit at fixed temperature is determined by the leading
eigenvalue $\lambda_1$, and this eigenvalue is common to the transfer
matrices $\mathbf{P}$ and $\mathbf{Q}$.
The leading-order correction to the magnetization, however, depends on
the eigenvalue with the second largest absolute value.
If we call this eigenvalue $\lambda_s$, then the average magnetization
for a chain of length $N$ behaves asymptotically as
\begin{eqnarray}
  \langle m\rangle = s_1 +
  \left(\frac{\lambda_s}{\lambda_1}\right)^{N-1}
  \frac{\partial}{\partial x}\left(\frac{\lambda_s}{\lambda_1}\right)
  + \text{higher order terms},
\end{eqnarray}
obtained by expanding the logarithm in Eq.~\ref{m_from_Z} and
inserting the definition of $s_1$ from Eq.~\ref{s1}.
In the unconstrained case we have $\lambda_s = \lambda_2$, but for the
constrained models one of the other two eigenvalues of the matrix
$\mathbf{Q}$ has a larger absolute value so long as $x\neq0$ or
$y\neq0$.
For example, if $x$ and $y$ are both positive, then
$\lambda_s=\lambda_4$ and the leading-order corrections for the
unconstrained, even and odd models are
\begin{eqnarray}
  &\langle m\rangle_{u} - s_1 =
    -2s_1\left(\frac{\lambda_2}{\lambda_1}\right)^N,
    \label{mu_lead_ord}\\
  &\langle m\rangle_{e} - s_1 = -\langle m\rangle_{o,N} + s_1 =
    (c_1-s_1)\left(\frac{\lambda_4}{\lambda_1}\right)^N,
    \label{meo_lead_ord}
\end{eqnarray}
respectively (see Eq.~\ref{c1} for the definition of $c_1$).
In general, $|\lambda_4|$ is considerably larger than $|\lambda_2|$.
As a consequence, the leading-order correction decays much more slowly in
the constrained cases than in the unconstrained one
(Fig.~\ref{lead_ord_corr}).

The difference in the asymptotic approach to the thermodynamic limit
becomes even more apparent if we take the limit $N\to\infty$ while
simultaneously $J\to\infty$ (so that $T_r\to 0$) and $H\to
0$ (so that $x\to 0$) in such a way that the products 
\begin{eqnarray}
  t\equiv NT_r,\label{t}\\
  h\equiv N\sinh x\label{h}
\end{eqnarray}
are constants.
In the thermodynamic limit the magnetic field $H$ scales $\propto N^{-1}$.
We could have alternatively defined $h=Nx$ to make this
inverse proportionality more apparent, but the definition of Eq.~\ref{h}
is a little bit more convenient when substituting the hyperbolic
functions in
Eq.~\ref{lambda_12} and \ref{lambda_34}.
After applying the formula $\lim_{N\to\infty}(1+z/N)^N=e^z$ to
Eq.~\ref{Z_e} and Eq.~\ref{Z_o}, we
obtain the partition functions for large $N$,
\begin{eqnarray}
  Z_u = 2N^{N/2}t^{-N/2}\cosh\sqrt{h^2+t^2},  \label{Zuht}
\\
  Z_{e,o} =
  \begin{cases}
  N^{N/2}t^{-N/2}\left(\cosh\sqrt{h^2+t^2}\pm\cosh h\right) 
& \text{if $N$ is even},\\
  N^{N/2}t^{-N/2}\left(\cosh\sqrt{h^2+t^2}\mp\sinh h\right)
& \text{if $N$ is odd}.
  \end{cases}
  \label{Zeoht}
\end{eqnarray}
All thermodynamic quantities can now be derived from the partition
function by taking the appropriate derivatives, for example $\langle
m\rangle = \partial(\ln Z)/\partial h$.
Alternatively, we can also take the thermodynamic limits of
Eq.~\ref{m_eo}, \ref{chieo_x0} and \ref{Ceo_x0}.
We tabulate the results in Table~\ref{scaling_functions}.

\bgroup
\def\arraystretch{1.5}
\setlength{\extrarowheight}{4pt}
\begin{table*}[!t]
\begin{tabular}{|l|c|c|c|}
  \hline
  & unconstrained & even model  &odd model\\
  \hline
  \multirow{2}{*}{$\langle m\rangle$}
  & \multirow{2}{*}{$\frac{h\tanh\sqrt{h^2+t^2}}{\sqrt{h^2+t^2}}$}
                  & $\frac{\frac
                    h{\sqrt{h^2+t^2}}\sinh\sqrt{h^2+t^2}+\sinh h}
                    {\cosh\sqrt{h^2+t^2}+\cosh h}$
                  &$\frac{\frac
                    h{\sqrt{h^2+t^2}}
                    \sinh\left(\sqrt{h^2+t^2}\right)-\sinh 
                    h}{\cosh\left(\sqrt{h^2+t^2}\right)-\cosh h}$
  if $N$ is even,\\
  & & $\frac{\frac
      h{\sqrt{h^2+t^2}}\sinh\sqrt{h^2+t^2}-\cosh h}
      {\cosh\sqrt{h^2+t^2}-\sinh h}$\rule{0pt}{5ex}
  & $\frac{\frac
  h{\sqrt{h^2+t^2}}\sinh\left(\sqrt{h^2+t^2}\right)+\cosh
  h}{\cosh\left(\sqrt{h^2+t^2}\right)+\sinh h}$ if $N$ is odd.\\
  \hline
  \multirow{2}{*}{$\frac{\chi(h=0)}{\beta}$}
  & \multirow{2}{*}{$\frac{N\tanh t}t$} 
    & $N\left(\frac{\tanh\left(\frac t2\right)}t +
      \frac1{2\cosh^2\left(\frac t2\right)}\right)$
                                &$N\left(\frac{\coth\left(\frac t2\right)}{t} -
    \frac1{2\sinh^2\left(\frac t2\right)}\right)$   if $N$ is even,\\
  & & \multicolumn{2}{c|}{$N\left(\frac{\tanh t}t-\frac1{\cosh^2t}\right)$  if $N$ is odd.}
    \rule{0pt}{3ex}\\
  \hline
  \multirow{2}{*}{$\frac{C(h=0)}{k_B\beta^2J^2}$}
    & \multirow{2}{*}{$\frac{4t}N\left(\tanh t + \frac t{\cosh^2
      t}\right)$}
                    & $\frac{2t}N\left(2\tanh\left(\frac t2\right) +
  \frac t{\cosh^2\left(\frac t2\right)}\right)$
  &$\frac{2t}N\left(2\coth\left(\frac t 2\right)-\frac t{\sinh^2\left(\frac t
  2\right)}\right)$ if $N$ is even\\
  & & \multicolumn{2}{c|}{$\frac{4t}N\left(\tanh t + \frac t{\cosh^2
      t}\right)$\rule{0pt}{3ex}  if $N$ is odd.}\\
  \hline
\end{tabular}
\caption{\label{scaling_functions} Thermodynamic properties for $N\to\infty$ and constant $t$,
  $h$ (defined in Eq.~\ref{t} and \ref{h}).}
\end{table*}

It is instructive to compare these equations with the canonical
finite-size scaling forms, for example for the susceptibility
\begin{eqnarray}
  \chi(h=0) &\propto N^{\gamma/\nu}f_\chi(N^{1/\nu}t).
\end{eqnarray}
While we find that $\gamma=\nu=1$ for all of the cases
listed in Table~\ref{scaling_functions} (see also our remark after
Eq.~\ref{crit_exp_nu}), the scaling functions $f_\chi$ (plotted in
Fig.~\ref{scal_func}) are fundamentally different.

\begin{figure*}
  \begin{center}
    \includegraphics[width=7.75cm]{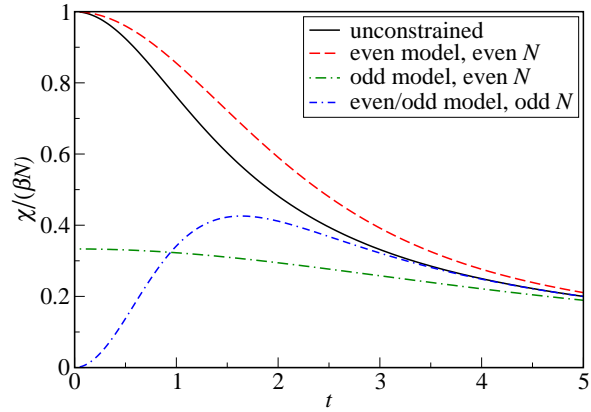}
    \caption{
      The scaling function of the susceptibility $\chi(h=0)$ in the limit
      $N\to\infty$, $\beta\to\infty$ with finite $t=Ne^{-2\beta J}$.
      The scaling functions for the unconstrained, even
      and odd models exhibit different behaviour, especially if $t$ is
      small. 
    }
    \label{scal_func}
  \end{center}
\end{figure*}

\section{The probability distribution of the magnetization}
\label{pdf_section}
Because of the differences between the unconstrained, even and odd
models in Table~\ref{scaling_functions}, one may wonder how the
probability distribution of the magnetization~\cite{Bruce81, Bruce85,
  Zheng03, Garcia09} differs;
after all,
$\langle m\rangle$ and $\chi$ are essentially the mean and variance of
this distribution.
We repeat here the arguments developed by Antal
et~al.~\cite{Antal_etal04} for the unconstrained model with zero
magnetic field.
We denote the total magnetization by $M\equiv\sum_i\sigma_i$ and the number
of domain walls (i.e.\ boundaries between stretches of contiguous
positive and negative spins) by $2d$; it must be an even number
because of the periodic boundary conditions.
The main task is to count the number $\Omega(d,M)$ of
configurations with $2d$ domain walls and magnetization $M$.
Their probability $P(d,M)$ in thermal equilibrium with $H=0$ will then
follow from
\begin{eqnarray}
  P(d,M) = \frac{e^{(N-4d)y}}{Z}\Omega(d,M),
  \label{PdM}
\end{eqnarray}
whose marginal distribution
\begin{eqnarray}
  P(M) = \sum_dP(d,M)
  \label{PM}
\end{eqnarray}
is the probability distribution we are looking for.

We can find $\Omega(d,M)$ with the following combinatorial argument.
Let us assume that there are $N_+$ positive and $N_-=N-N_+$ negative
spins, and that the first spin is positive.
We could for example have 
\begin{eqnarray}
  \underbrace{w_0+}++\underbrace{w_1-}---\underbrace{w_2+}++\ldots++\underbrace{w_{2d-1}-}w_{2d}+++,
  \label{domain_wall}
\end{eqnarray}
where we marked the positions of the domain walls by $w_1,\ldots,
w_{2d}$.
At the periodic boundary between the first and last spin there may not
be a domain wall (in the example above there is not), but we will
always symbolically put $w_0$ in front of the chain.
We now mentally glue $w_0,\ldots, w_{2d-1}$ to the next spin in the chain
(indicated by the braces in Eq.~\ref{domain_wall}).
In this manner, $d$ negative spins are attached to domain walls,
whereas the remaining $N_--d$ can be freely placed in the $d$
negative domains.
The well-known stars-and-bars theorem~\cite{Feller50} implies that
there are $\bigl(\begin{smallmatrix}N_--1\\d-1\end{smallmatrix}
\bigr)$ different ways to distribute the negative spins.

The positive spins require a little more care, because there may not be a
positive spin trailing the domain wall $w_{2d}$.
We can account for this exception by not attaching $w_{2d}$ to the
following spin.
There are thus $d$ positive spins attached to $w_0,w_2,\ldots,
w_{2d-2}$, while the remaining $N_+-d$ positive spins can be freely distributed
into $d+1$ segments, namely the positive intervals following
$w_0,w_2,\ldots, w_{2d}$.
According to the stars-and-bars theorem, there are
$\bigl(\begin{smallmatrix}N_+\\d\end{smallmatrix} \bigr)$ different
possibilities.

Because the positive and negative spins are placed independently of
each other, the number of configurations is simply the product of the
binomial coefficients $\bigl(\begin{smallmatrix}N_--1\\d-1\end{smallmatrix}
\bigr) \bigl(\begin{smallmatrix}N_+\\d\end{smallmatrix} \bigr)$.
If we had started the chain with a negative spin, we would have
obtained the same expression with the subscripts $+$ and $-$
interchanged, so that
\begin{eqnarray}
  \Omega(d,M) = \binom{N_--1}{d-1}\binom{N_+}d +
  \binom{N_+-1}{d-1}\binom{N_-}d.
\end{eqnarray}

This expression from Antal~et~al.~\cite{Antal_etal04} is equally valid
for the unconstrained, even and odd model. The constraints only enter
in the permitted values for $M$ whose consequence becomes apparent
when we take the continuum limit.
To this end, we take $N\to\infty$ for a fixed value of $d$ and write
$m=M/N$, so that
\begin{eqnarray}
  \Omega(d,m) =
  \frac{N^{2d-1}}{2^{2d-2}\,d!\,(d-1)!}\left(1-m^2\right)^{d-1}.
  \label{Omega_dm}
\end{eqnarray}
For the time being let us assume that $|m|\neq1$ and thus $d\neq0$.
We can insert Eq.~\ref{Omega_dm} into Eq.~\ref{PdM} and \ref{PM}, but
have to bear in mind that changing from the discrete variable $M$ to the
continuous variable $m$ generates an additional prefactor, which we
will call $N/\Delta_M$,
\begin{eqnarray}
  P(m) =
  \frac{4N^{N/2}}{t^{N/2}\Delta_MZ(1-m^2)}\sum_{d=1}^\infty\frac{t^{2d}(1-m^2)^d}{2^{2d}d!(d-1)!}.
  \label{Pdm}
\end{eqnarray}
Here $\Delta_M$ is the step size between consecutive values of $M$ (i.e.\
$\Delta_M=2$ in the unconstrained, $\Delta_M=4$ in the even and odd
model) and $t$ is defined in Eq.~\ref{t}.
As noticed in Ref.~\cite{Antal_etal04}, the infinite series in
Eq.~\ref{Pdm} can be expressed in terms of a modified Bessel function
of the first kind thanks to the identity~\cite{AbramowitzStegun72}
\begin{eqnarray}
  I_1(z) = \sum_{d=0}^\infty \frac{z^{2d+1}}{2^{2d+1}d!(d+1)!}
\end{eqnarray}
and therefore
\begin{eqnarray}
  P(m) =
  \frac{2N^{N/2}I_1\left(t\sqrt{1-m^2}\right)}
  {t^{N/2-1}\Delta_MZ\sqrt{1-m^2}}
  \label{Pm}
\end{eqnarray}
for $|m|<1$.

At the boundaries of this interval (i.e.\ $|m|=1$) there are
contributions proportional to Dirac delta functions.
These singularities arise because $|m|=1$ implies $d=0$, leaving the
denominator in Eq.~\ref{Omega_dm} undetermined.
For the unconstrained model as well as the even and odd model with
even $N$, the proportionality constants in front of the delta
functions can be computed based on the observation that $P(m)$ must be
normalized and symmetric about $m=0$.
For the even model with odd $N$, a discrete magnetization $M=-N$ is
permitted, but $M=N$ is not, so that a delta function can only appear
at $m=-1$, but not $m=1$.
Conversely, the odd model with odd $N$ can only have a singular
contribution at $m=1$, but not at $m=-1$.

The probability contained in the regular part of the distribution
given by Eq.~\ref{Pm} follows from the integral
\begin{eqnarray}
  \int_{-1}^1\frac{I_1\left(t\sqrt{1-m^2}\right)}{\sqrt{1-m^2}}\,dm =
  \frac{4\sinh^2(t/2)}t
\end{eqnarray}
and, upon inserting the partition functions of Eq.~\ref{Zuht} and
\ref{Zeoht} with $h=0$, we obtain
\begin{eqnarray}
  P_u(m) = \frac{tI_1\left(t\sqrt{1-m^2}\right)}{2\sqrt{1-m^2}\cosh t}
  + \frac{\delta(m-1)+\delta(m+1)}{\cosh t},
  \label{P_um}\\
  P_e(m) = 
  \begin{cases}
    \frac{tI_1\left(t\sqrt{1-m^2}\right)}{4\sqrt{1-m^2}\cosh^2(t/2)} +
    \frac{\delta(m-1)+\delta(m+1)}{\cosh^2(t/2)} &\text{if $N$ is
      even,}\\
    \frac{tI_1\left(t\sqrt{1-m^2}\right)}{2\sqrt{1-m^2}\cosh t} +
    \frac{2\delta(m+1)}{\cosh t} &\text{if $N$ is odd},
  \end{cases}
  \label{P_em}\\
  P_o(m) =
  \begin{cases}
    \frac{tI_1\left(t\sqrt{1-m^2}\right)}{4\sqrt{1-m^2}\sinh^2(t/2)}
    &\text{if $N$ is even},\\
    \frac{tI_1\left(t\sqrt{1-m^2}\right)}{2\sqrt{1-m^2}\cosh t} +
    \frac{2\delta(m-1)}{\cosh t} &\text{if $N$ is odd}.
  \end{cases}
                                   \label{P_om}
\end{eqnarray}

One noteworthy detail is that the delta functions peak exactly at the
boundaries of the interval $[-1,1]$.
So long as the integral of the delta function over the entire real
line equals $1$, it is a matter of definition how much weight is
assigned to the left and right of the interval boundaries.
We have adopted here the symmetric convention $\int_0^\infty\delta(x)dx
= 1/2$ which applies, for instance, if the delta function is the limit
of narrowing zero-centred Gaussians.
Other conventions are possible; for example Ref.~\cite{Antal_etal04}
implicitly uses $\int_0^\infty\delta(x)dx=1$ which changes the
prefactors in front of the delta functions in
Eq.~\ref{P_um}--\ref{P_om}.
With our definition of the delta function and the integral
\begin{eqnarray}
  \int_{-1}^1\frac{m^2I_1\left(t\sqrt{1-m^2}\right)}{\sqrt{1-m^2}}dm = 
  \frac2t\left(\frac{\sinh t}t-1\right), 
\end{eqnarray}
we can indeed retrieve the susceptibility $\chi$ in
Table~\ref{scaling_functions}.

\section{Discussion}
\label{discussion}

\begin{figure*}
  \begin{center}
    \includegraphics[width=15.5cm]{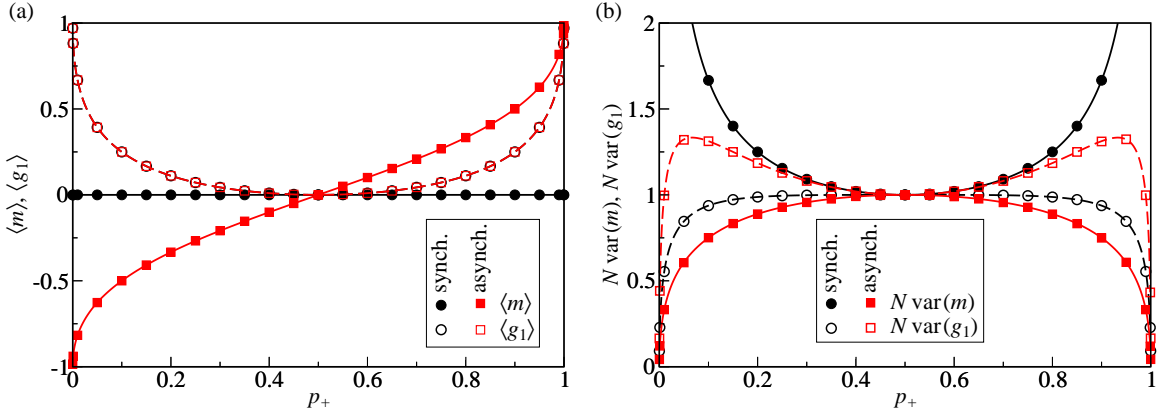}
    \caption{
      (a) Mean nearest-neighbour correlation $\langle m\rangle$
      and second-nearest neighbour correlation $\langle g_1\rangle$ in
      the stochastic voter model.
      Black curves and symbols are for synchronous, red for
      asynchronous updates.
      Analytic predictions (Eq.~\ref{m_synch}, \ref{g1_synch},
      \ref{m_asynch}, \ref{g1_asynch}) are shown as
      solid and dashed curves.
      The results of Monte Carlo simulations for a chain of length
      $N=100$ are shown as circles and squares.
      (b) The same for the variances of $m$ and $g_1$
      (Eq.~\ref{varm_synch}, \ref{varg1_synch}, \ref{varm_asynch},
      \ref{varg1_asynch}).
    }
    \label{discuss_fig}
  \end{center}
\end{figure*}

Combining the results above, we can now analytically solve the
stochastic synchronous and asynchronous voter models introduced in
Sec.~\ref{motivation} and \ref{asynch_section}.
With Eq.~\ref{voter_sigma} we can translate $m$ and $g_1$ of the Ising
model into correlations between the opinions of nearest and
next-nearest neighbours,
\begin{eqnarray}
  &\langle m\rangle =
  \frac1N\left\langle\sum_i\omega_i\omega_{i+1}\right\rangle,\\
  &\langle g_1\rangle =
  \frac1N\left\langle\sum_i\omega_i\omega_{i+2}\right\rangle,
\end{eqnarray}
where the second equation follows from Eq.~\ref{g1_from_Z} and
$\omega_{i+1}^2=1$.
The variances of $m$ and $g_1$ are proportional to the second partial
derivatives of $\ln Z$ with respect to either $x$ or $y$, thus
\begin{eqnarray}
  &N\var(m) = \frac{\chi}{\beta},\\
  &N\var(g_1) = \frac1N\frac{\partial^2}{\partial y^2}\ln Z,
\end{eqnarray}
where the derivative in the last equation has to be evaluated at $x=0$
for synchronous and $y=0$ for asynchronous updates.
For synchronous updates, we have in fact evaluated this derivative
already in Eq.~\ref{Ceo_x0} because in this case $N\var(g_1) =
C/(k_B\beta^2J^2)$. 
The corresponding calculation for asynchronous updates can be
performed by differentiating the partition functions in Eq.~\ref{Z_e}
and \ref{Z_o}.

Inserting Eq.~\ref{voter_beta} into Eq.~\ref{meo_x0}, \ref{chieo_x0},
\ref{g1eo_x0} and \ref{Ceo_x0}, we obtain for the synchronous voter model
in the thermodynamic limit
\begin{eqnarray}
  \lim_{N\to\infty}\langle m\rangle =
  \begin{cases}
    1 & \text{if $N$ is odd and either $p_+=0$ or $p_+=1$},\\
    0 & \text{otherwise},
  \end{cases}
        \label{m_synch}
\end{eqnarray}
\begin{eqnarray}
  \lim_{N\to\infty}[N\var(m)] = \frac1{2\sqrt{p_+p_-}},
  \label{varm_synch}
\end{eqnarray}
\begin{eqnarray}
  \lim_{N\to\infty}\langle g_1\rangle = \frac{1-2\sqrt{p_+p_-}}
  {1+2\sqrt{p_+p_-}},
  \label{g1_synch}
\end{eqnarray}
\begin{eqnarray}
  \lim_{N\to\infty}[N\var(g_1)] =
  \frac{8\sqrt{p_+p_-}}{(1+2\sqrt{p_+p_-})^2},
  \label{thermolim_g1}
  \label{varg1_synch}
\end{eqnarray}
where, as before, $p_-=1-p_+$.
For asynchronous updates the corresponding results are
\begin{eqnarray}
  \lim_{N\to\infty}\langle m\rangle =
  \frac{\sqrt{p_+}-\sqrt{p_-}}{\sqrt{p_+}+\sqrt{p_-}},
  \label{m_asynch}\\
  \lim_{N\to\infty}[N\var(m)] =
  \frac{4\sqrt{p_+p_-}}{1+2\sqrt{p_+p_-}},
  \label{varm_asynch}\\
  \lim_{N\to\infty}\langle g_1\rangle = \frac{1-2\sqrt{p_+p_-}}
  {1+2\sqrt{p_+p_-}},
  \label{g1_asynch}\\
  \lim_{N\to\infty}\left[N\var(g_1)\right] =
  \frac{16\sqrt{p_+p_-}(1-\sqrt{p_+p_-})}
  {1+4\sqrt{p_+p_-}(1+\sqrt{p_+p_-})}.
  \label{varg1_asynch}
\end{eqnarray}
We plot Eq.~\ref{m_synch}--\ref{varg1_asynch} in
Fig.~\ref{discuss_fig}.
As a numerical confirmation we include the results of Monte Carlo
simulations in the same graphs.
The numerical and analytic results are in excellent agreement.

Comparing the synchronous with the asynchronous case, we notice that
the thermodynamic limits of the nearest-neighbour correlations
$\langle m\rangle$ differ significantly.
While for synchronous updates nearest neighbours are typically
uncorrelated, asynchronous updates build up non-zero correlations.
Interestingly, the mean second-nearest neighbour correlations $\langle
g_1\rangle$ are identical for both update rules.
However, the variances differ between the rules: $\var(m)$ is larger
for synchronous updates, whereas $\var(g_1)$ is larger for
asynchronous updates.

It is in principle possible to extend the calculations to correlations
between more distant neighbours too.
For example, $\left\langle\sum\omega_i\omega_{i+3}\right\rangle$ can
be obtained by formally introducing a three-spin interaction strength
$K$ in the Hamiltonian so that $E(\boldsymbol{\sigma}) =
-K\sum_i\sigma_i\sigma_{i+1}\sigma_{i+2} - J\sum_i\sigma_i\sigma_{i+1}
- H\sum\sigma_i$.
The mean third-nearest neighbour correlation follows from
differentiating the partition function $Z$ with respect to $K$ and
subsequently setting $K=0$ as well as $H=0$ for synchronous, $J=0$ for
asynchronous updates.
Unfortunately, the transfer matrix method developed in
Sec.~\ref{partition_function} does not easily generalize to arbitrary
$k$-spin interactions~\cite{Fan11}, but calculating correlations in
the voter model from Ising-like Hamiltonians is an intriguing
possibility for future research.

\section{Conclusion}
\label{conclusion}
We have studied two variants of the one-dimensional Ising model: in
the first variant the number of positive spins is constrained to an
even number; in the second model this number must be odd.
We have motivated both models by mapping them to a model of opinion
dynamics with either synchronous or asynchronous updates.
If the temperature and magnetic field are held constant, the
thermodynamic limits of the even and odd Ising models are the same as
the limit of the unconstrained model.
However, by simultaneously increasing the chain length and
lowering the temperature and magnetic field, we have shown that the
scaling functions for the even, odd and unconstrained models differ.
The mapping from the Ising model has allowed us to obtain explicit
formulae for correlations between nearest and next-nearest neighbours
in the voter model.

We can generalize the problem posed in this paper to higher dimensions
or complex networks by associating spins with the links and
enforce an even number of positive spins on every cycle in the graph.
In other words, only balanced signed graphs~\cite{Harary53} are
permitted.
Assigning opinions to the nodes and mapping them to spins on the links
as in Eq.~\ref{voter_sigma} will naturally generate such graphs from
the voter model.
It is a fascinating question how this changes the thermodynamic limits
compared to the unconstrained model.

\ack
This research is supported by the
European Commission (project number FP7-PEOPLE-2012-IEF 6-4564/2013).
We thank Zolt\'an R\'acz for helpful discussions.\\

\end{document}